\documentclass[sigconf]{acmart}
\usepackage{svg}

\renewcommand\footnotetextcopyrightpermission[1]{} 

\AtBeginDocument{%
  \providecommand\BibTeX{{%
    \normalfont B\kern-0.5em{\scshape i\kern-0.25em b}\kern-0.8em\TeX}}}

\setcopyright{acmcopyright}
\copyrightyear{2022}
\acmYear{2022}
\acmDOI{10.1145/3494106.3528680}

\acmConference[BSCI '22]{Proceedings of the Fourth ACM International Symposium on Blockchain and Secure Critical Infrastructure }{May 30, 2022}{Nagasaki, Japan}
\acmBooktitle{Proceedings of the Fourth ACM International Symposium on Blockchain and Secure Critical Infrastructure (BSCI '22), May 30, 2022, Nagasaki, Japan}
\acmPrice{15.00}
\acmISBN{978-1-4503-9175-7/22/05}

\begin{document}
\fancyhead{}
\title{ScaleSFL:\\ A Sharding Solution for Blockchain-Based Federated Learning}


\author[1]{Evan Madill}
\orcid{0000-0002-6668-3671}
\affiliation{%
  \department{Department of Computer Science}
  \institution{University of Manitoba}
  \city{Winnipeg}
  \state{MB}
  \country{Canada}
}
\email{madille@myumanitoba.ca}

\author[2]{Ben Nguyen}
\affiliation{%
  \department{Department of Computer Science}
  \institution{University of Manitoba}
  \city{Winnipeg}
  \state{MB}
  \country{Canada}
}
\email{nguyen53@myumanitoba.ca}

\author[3]{Carson K. Leung}
\orcid{0000-0002-7541-9127}
\affiliation{%
  \department{Department of Computer Science}
  \institution{University of Manitoba}
  \city{Winnipeg}
  \state{MB}
  \country{Canada}
}
\email{kleung@cs.umanitoba.ca}
\email{Carson.Leung@UManitoba.ca}

\author[4]{Sara Rouhani}
\affiliation{%
  \department{Department of Computer Science}
  \institution{University of Manitoba}
  \city{Winnipeg}
  \state{MB}
  \country{Canada}
}
\email{sara.rouhani@umanitoba.ca}
\authornote{Corresponding authors}

\renewcommand{\shortauthors}{E. Madill et al.}

\begin{abstract}
    Blockchain-based federated learning has gained significant interest over the last few years with the increasing concern for data privacy, advances in machine learning, and blockchain innovation. However, gaps in security and scalability hinder the development of real-world applications. In this study, we propose ScaleSFL, which is a scalable blockchain-based sharding solution for federated learning. ScaleSFL supports interoperability by separating the off-chain federated learning component in order to verify model updates instead of controlling the entire federated learning flow. We implemented ScaleSFL as a proof-of-concept prototype system using Hyperledger Fabric to demonstrate the feasibility of the solution. We present a performance evaluation of results collected through Hyperledger Caliper benchmarking tools conducted on model creation. Our evaluation results show that sharding can improve validation performance linearly while remaining efficient and secure.
\end{abstract}




\maketitle

\section{Introduction}
With the increasing concern for data privacy and security of personal data, the demand for creating trusted methods of privacy-preserving communication is rapidly increasing. Recent advances in machine learning have also spurred interest in a variety of privacy-sensitive domains such as healthcare informatics \cite{ref:healthcare-fl} and coronavirus disease 2019 (COVID-19) diagnoses \cite{ref:covid-19-diagnosis-fl}. However, the privacy of personal data used to train these models has been called into question. Traditional methods of training require data to be centrally located. However, this is challenging or impossible with privacy regulations.

Federated learning (FL) \cite{ref:federated-learning} is a proposed solution allowing sensitive data to remain distributed among edge devices. This framework will enable devices to train models locally on-device and then upload their local model updates to a trusted central server to be aggregated, resulting in a global model that is then distributed. This process is repeated until the desired objective is reached, such as the convergence of model accuracy.

Additional improvements have been made to the aggregation strategies involved in traditional FL, notably in reducing the communication overhead to the central aggregation server. Solutions such as HierFavg \cite{ref:hierfavg} introducing a hierarchical scaling solution, where model updates are instead sent to local edge servers performing partial model aggregation. This combines the benefits of both centralized cloud and edge computing, resulting in both reduced latency and communication overhead. 

While the traditional FL framework has demonstrated its viability in real-world applications, it has several notable issues. First, model updates can indirectly leak the local data used to train them. Specifically, users can be de-anonymized through data reconstruction via statistical analysis and data mining. An example of this was with the Netflix Prize Dataset \cite{ref:deanonymize-netflix} where data was correlated to users using an auxiliary IMDb dataset. The second issue is the centralized nature of aggregating the model updates, which creates a single point of failure and relies on trusted authorities.
    
Differential privacy \cite{ref:differential-privacy, ref:differential-privacy-impossibility}, a mathematical framework for evaluating dataset privacy of a given model, can be employed to solve the first issue, for example, by adding a certain noise to the model update. A solution to the second issue is to utilize blockchain technologies to decentralize the task of aggregating the local model updates. However, using a decentralized system incurs several more issues by removing the presence of a centrally trusted entity which gives rise to, for example, the problem of detecting malicious clients sending false model updates.
    
Additionally, because local datasets do not necessarily represent the global distribution, as they are non-IID (independent and identically distributed), two honest clients could submit opposing model updates. This problem can be addressed by validating each transaction. One possible solution is to allow each node to execute validation procedures in a smart contract, as described in BAFFLE \cite{ref:baffle}. However, delegating this task to every node of the network is computationally expensive considering the high overhead, local model training, and the consensus task \cite{ref:committee-consensus}. Another solution is to select a committee of nodes each round to perform this validation task \cite{ref:committee-consensus}.

We can define the current number of clients training each round as $\mathcal{C}$, and the number of committee nodes (endorsing peers) as $\mathcal{P}_E$. In each round, a total of $\mathcal{C} \times \mathcal{P}_E$ computations must be completed to validate all transactions. 

Here, we propose a sharding mechanism extending the committee consensus method. As the network expands, clients will be assigned to different shards within the network, and a committee will be elected for each shard. After every phase, each shard's committee will publish a local shard-aggregated model, which will then be globally aggregated and redistributed to all clients. If we have $\mathcal{S}$ many shards, this would reduce the global computations from $\mathcal{C} \times \mathcal{P}_E$ to $\frac{\mathcal{C} \times \mathcal{P}_E}{\mathcal{S}^2}$ for each shard, or $\frac{\mathcal{C} \times \mathcal{P}_E}{\mathcal{S}}$ globally.


To preserve the security of the system across shards, we propose a flexible protocol allowing for a pluggable poisoning mitigation, and detection strategies \cite{ref:roni, ref:krum, ref:fools-gold, ref:attack-adaptive-aggregation}. This allows proposed models to adopt suitable strategies, while allowing the system evolves with recent advances.

Our {\em key contributions} of this paper can be summarized as follows:
\begin{itemize} 
    \item a \textbf{sharding framework} called ScaleSFL, which supports a sharding solution for blockchain-based federated learning. We formally discuss the design of our algorithm for the sharding mechanism based on two consensus levels: shard-level consensus and mainchain consensus.
    \item an \textbf{implementation} of a proof-of-concept prototype system using Hyperledger Fabric \cite{ref:fabric} to run local experiments and evaluate the solution effectiveness.
    \item \textbf{benchmarks} that we demonstrate the performance of our implementation by using Hyperledger Caliper benchmarking tools \cite{ref:hyperledger-caliper}, showing our framework scales linearly with the number of shards.
\end{itemize}

The remainder of this paper is organized as follows. The next section provides background and related works. Our ScaleSFL framework, a sharding solution for blockchain-based federated learning, is introduced in Section~\ref{sec3}. Evaluation and discussion are presented in Sections~\ref{sec4} and \ref{sec5}, respectively. Finally, we conclude our work, and describe future works in Section~\ref{sec6}. 

\section{Background and Related Works}
Our current work integrates the ideas of federated learning and differential privacy, as well as blockchain-based federated learning. In this section, we introduce individual topics and corresponding hybrid solutions to provide background.

\subsection{Federated Learning and Differential Privacy}
Privacy-preserving machine learning combines the use of federated learning with differential privacy \cite{ref:privacy-preserving-deep-learning, ref:differential-privacy-sgd, ref:cit21}. Federated learning allows sensitive data from edge devices (e.g., smartphones, tablets, IoT devices) to remain on-device while training is done locally (distributed training on edge devices) \cite{ref:federated-learning}. In the original FL environment, a centralized server manages the training process. The server broadcasts the current model during each round to a subset of participating nodes to activate local training (requesting a model update). After all model update submission responses are received, the server aggregates the local models into a global model and prepares for the next round \cite{ref:federated-learning, ref:committee-consensus}. While FL offers many privacy benefits, differential privacy in conjunction provides stronger privacy guarantees \cite{ref:federated-learning} by mitigating data reconstruct-ability via random model update noise.

McMahan et al. \cite{ref:federated-learning} proposed the Federated Averaging algorithm (FedAvg). Their algorithm is a generalization of federated stochastic gradient descent (FedSGD) \cite{ref:differential-privacy-sgd} allowing each node to perform multiple batch updates while exchanging the models' parameters, with which we based our FL model training approach. Experiments demonstrate that FL can be used for training a variety of high-quality machine learning models at reasonable communication cost, exemplifying high practicality \cite{ref:federated-learning}. Therefore, we choose to apply differential privacy and use an adapted version of FedAvg for our current work.

Liu et al. \cite{ref:hierfavg} introduced HierFavg, a hierarchical scaling solution introducing multiple edge servers, each performing partial model aggregation, combining the benefits of both centralized cloud and edge computing. While centralized servers benefit from the vast reach of data collection, the communication overhead and latency of downloading and evaluating such updates limit this approach. Instead, model updates are sent to local edge servers to benefit from fast and efficient model updates, reducing the number of local iterations and total runtime. These local updates are then sent to a cloud server allowing for a sampling of the global data distribution to be considered and outperforming the local edge server performance. Combining these approaches is nontrivial, and the authors performed convergence analysis describing qualitative guidelines in applying their approach. While our approach is primarily motivated by increasing the throughput of model update metadata to blockchain ledgers, the composition of shards can benefit from the strategies found in hierarchical solutions.

\subsection{Blockchain-Based Federated Learning}
Blockchain technology can be utilized to decentralize the aforementioned centralized server in federated learning. The aggregation and facilitation logic previously done by the central server can be replaced by implementing smart contracts \cite{ref:committee-consensus}.

In addition to Ramanan and Nakayama's blockchain-based aggregator free federated learning (BAFFLE) \cite{ref:baffle} and Li et al.'s blockchain-based federated learning framework with committee consensus \cite{ref:committee-consensus} referenced earlier, numerous strategies for integrating blockchain technology with federated learning to achieve decentralization have been considered \cite{SatybaldyN20}. However, the approaches proposed by previous works do not directly focus on the performance of their chosen underlying consensus mechanism with which their proposed frameworks are built upon. 

\subsubsection{Blockchain Integration}
Kim et al.'s blockchain-based federated learning (BlockFL) \cite{ref:bc-on-device-fl} was implemented using the proof-of-work consensus mechanism and focused on analyzing the end-to-end latency of their system. Ma et al.'s proposed blockchain-assisted decentralized FL (BLADE-FL) \cite{ref:fl-meets-bc} was able to address the single point of failure issue while analyzing the learning performance. Liu et al.'s proposed federated learning with asynchronous convergence (FedAC) method \cite{ref:bfl-async} seeks to improve communication performance through integrating asynchronous global aggregation. While these works focus on the performance of various system components, they do not consider the scalability requirements for large numbers of participants and the impact on the consensus mechanism's throughput.

Additionally, many blockchain-based federated learning applications with domain-specific frameworks have also been proposed. For example, healthcare \cite{ref:bfl-healthcare}, IoT \cite{ref:bfl-iot} and 6G networks \cite{ref:bfl-6g}. Similarly, these works focus on cutting-edge implementation plausibility rather than scalability or details regarding the performance of their consensus mechanisms.

Gai et al. \cite{ref:bfl-iiot} proposed a model that utilizes a blockchain-based solution to develop a privacy-preserving environment for task allocations in a cloud/edge system. Employing blockchain enhances trustworthiness in edge nodes and security in communications. Utilizing smart contracts also contributes toward reaching optimal task allocations. The model applies a differential privacy algorithm (Laplace distribution mechanism \cite{ref:differential-privacy}) to process energy cost data in edge nodes. 

Li et al. \cite{ref:committee-consensus} proposed a blockchain-based FL framework with committee consensus in response to consensus mechanism efficiency and framework scalability. The committee consensus mechanism works as follows: A subset of nodes constitutes a committee responsible for validating model updates and new block generation. Meanwhile, the other nodes only execute their local model training and send their updates to the committee nodes. This committee is re-elected every round based on scores from the previous round, or alternatively, re-election is randomized for implementation simplicity. Li et al. demonstrated that committee consensus is more efficient and scalable while remaining secure.

\subsubsection{Sharding Solutions}
Sharding allows for scalability by splitting the network into smaller subgroups. This reduces storage and communication requirements while increasing throughput as the number of shards increases.

Yuan et al. \cite{ref:bfl-shard} proposed ChainFL, a two-layered blockchain-driven FL system that splits an IoT network into multiple shards. Also using Hyperledger Fabric \cite{ref:fabric}, ChainFL adopts a direct acyclic graph (DAG)-based mainchain to achieve parallel (and asynchronous) cross-shard validation. While their approach is DAG-based, we focus on a hierarchical FL solution, and are able to use existing off-chain FL frameworks to increase interoperability.

\subsection{Model Update Poisoning}\label{sec:model-defence}
A notable issue when aggregating model updates from untrusted and potentially adversarial clients are model poisoning attacks. In order to disrupt or bias the training process, and adversary may submit biased, or corrupted model updates. Because model updates must be verified prior to being accepted to the ledger, we consider each defence mechanism applied to each endorsing peer rather than a single centralized server. One possibility to make the training more robust is to apply norm constraints to each model update, such as clipping model updates and applying random noise to the aggregated model to prevent overfitting \cite{ref:fl-survey}. However, further mechanisms are required to counter adversarial examples and backdoors \cite{ref:fl-backdoor}. A subset of model poisoning is data poisoning attacks. These occur when the clients' dataset has been altered, such as a targeted attack by adversarial manipulation of edge devices, or untargeted such as a faulty data collection process, resulting in corrupted samples (e.g., incorrect labels). Here, we focus on readily applicable works related to the general detection of poisoned model updates.

Barreno et al. \cite{ref:roni} proposed a technique to determine which samples from a training set are malicious. To do so, they present the reject on negative influence (RONI) defence. This method measures the effect each training sample has on the accuracy of the model. This defence strategy is appropriately modified to suit the needs of a FL setting. Instead, we measure the influence of each model update on the accuracy of the global model. This can be done by utilizing a held-out testing set, potentially unique to each endorsing peer. This defence is not suitable for general FL applications due to the nature of client data being non-IID and incorrectly rejected due to misrepresentation by the test set. Additionally, malicious updates may go unnoticed due depending on the threshold of acceptance.

Blanchard et al. \cite{ref:krum} introduced a method designed to counter adversarial updates in a FL setting. Multi-Krum is a byzantine-resilient defence strategy that selects $n$ vectors furthest from the calculated mean, removing them from the aggregated gradient update. These distances are calculated based on the euclidean distance between gradient vectors. This strategy is effective for attacks with up to 33\% compromised/adversarial clients. However, the calculated mean can be influenced by the Sybil updates and may fail for an attacker with more significant influence over the network. 

Fung et al. \cite{ref:fools-gold} proposed a novel defence strategy to discriminate between honest clients and Sybils based on the diversity of their respective gradient updates. The intuition behind this defence is to identify Sybils based on a shared objective, and thus their models' updates are likely to have low variance. This works by using a cosine similarity, looking for similarity in terms of indicative features (i.e., those that contribute to the correctness of the update). This approach can be further augmented with other defence methods such as Multi-Krum to determine the poisoned updates.

Wan and Chen \cite{ref:attack-adaptive-aggregation} introduced an attack-adaptive defence utilizing an attention-based neural network to learn plausible attacks. Instead of using similarity-based mechanisms, this introduces tailored defence strategies to prevent backdoor attacks. 

Ma et al. \cite{ref:dp-data-poisoning} explored differential privacy as a data poisoning defence based on creating a robust learning algorithm. While this defence is ideal for systems already using differential privacy, it enforces the use of noise in the learning algorithm, hurting performance. Additionally, this assumes a small subset of malicious clients, failing to defend against larger coordinated attacks.

\begin{figure*}[!th]
	\centering
 	\includegraphics[width=0.9\linewidth]{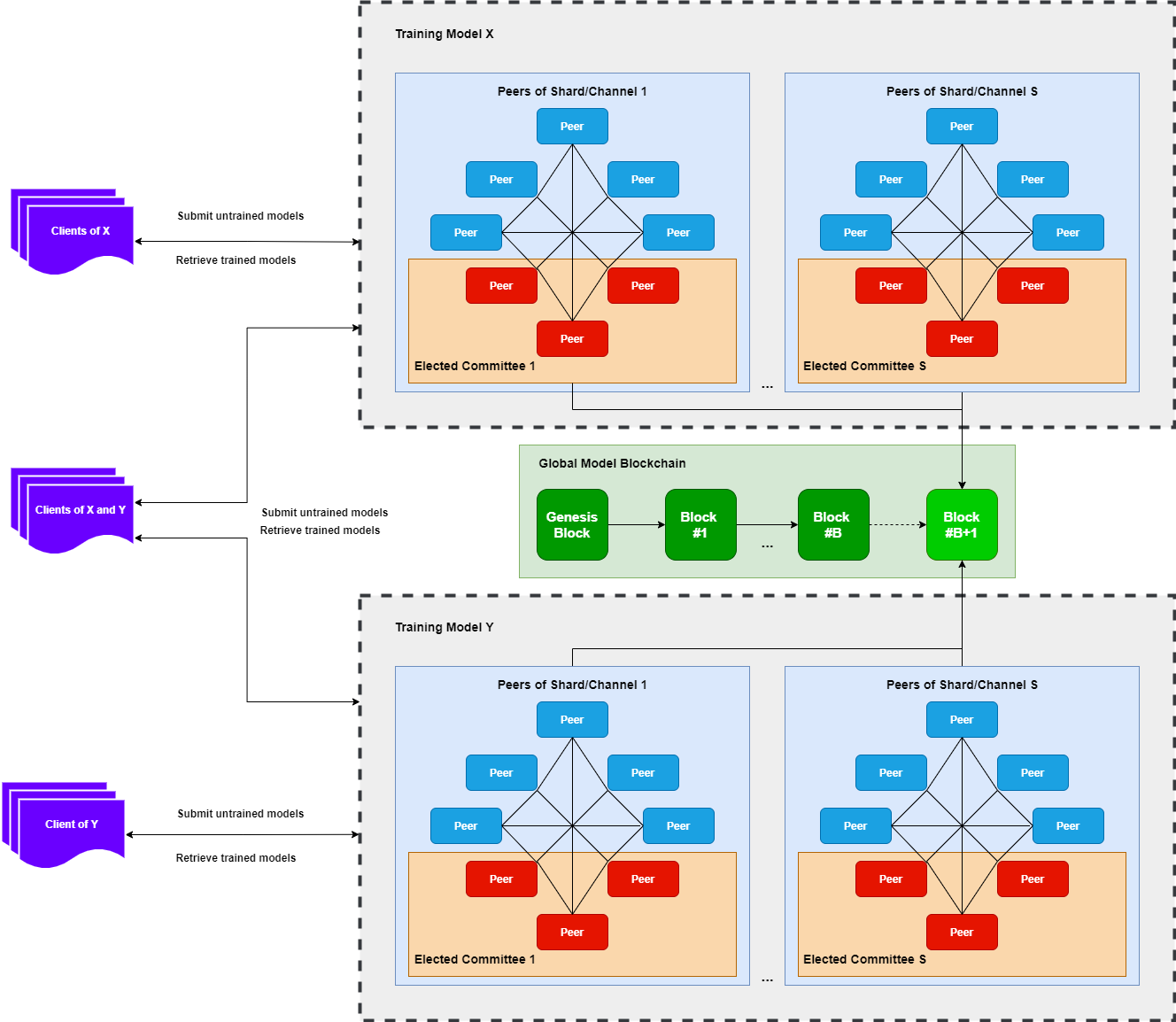}
	\caption{Global overview of our FL sharding approach}
	\label{fig:sharding-architecture}
\end{figure*}

Related attacks, such as model update plagiarism \cite{ref:fl-meets-bc}, may additionally occur, leading to a misleading distribution of data amongst clients. A malicious user may attempt to submit the same model update multiple times through several clients, known as lazy clients. They may then attempt to reap the rewards of contribution for each of these submissions. These can be detected by using a PN-sequence \cite{ref:fl-meets-bc, ref:blade-fl-analysis}, whereby the initial model update is submitted with random noise applied to the update and is then further verified by other clients using their own PN-sequence, checking for correlation. 



\section{Our Sharding Solution}\label{sec3}

This section describes our approach and outlines the architecture for our proposed sharding solution. An outline of the workflow can be seen in Figure~\ref{fig:sharding-architecture}. We can break this into two primary components, the mainchain, which coordinates aggregated models from shards, and the individual shard chains. 

Our framework follows a few core principles. Our solution is modular such that (a) the policies that govern model acceptance are abstracted, for example, model update poisoning, and mitigation strategies (Section~\ref{sec:model-defence}), are supported to determine model acceptance. Our framework additionally (b) supports external FL coordination, such that new paradigms such as data-centric FL are supported. This means the source of models updates is agnostic. 

With these principles, our solution acts as an additional component to existing FL workflows such that it handles model provenance, the evolution of security and poisoning defences, scalability through sharding, and the possibility of reward distribution for future works. Therefore, we solve the initial problem of FL systems being dependent on a central service to aggregate updates, providing a trusted system to pin verified model updates. We will further discuss the components of this system below.

\subsection{FL Algorithm}
The underlying problem of federated learning is to produce a global model from client updates trained on local privacy-sensitive datasets. We can define each client $\mathcal{C}_k$ to have their own set of data $D_k \subseteq D, \forall k, 0 \leq i \leq K$.  Using this, we can define the global objective to optimize as a modification of the FedAvg algorithm \cite{ref:federated-learning} as
\begin{equation} \label{eq:global-shard-objective-problem}
    \underset{w \in R^d}{min} f(w) \quad \textrm{and} \quad \underset{w \in R^d}{min} F_k(w) \quad \textrm{for each client}
\end{equation}
where the batch loss for each client is defined as 
\begin{equation}
    f_{b_{k,i}}(w) = \frac{1}{|b_k|} \sum_{x_i \in b_{k,i}} l(x_i, w_k)    
\end{equation}
and $b_{k,i}$ being a random subset from $D_k$. Each client updates their local objective with respect to their dataset $D_k$ by
\begin{equation}
    w_k \leftarrow w_k - \eta_k \nabla f_{b_{k,i}}(w_k)
\end{equation}
where $\eta_k$ is the clients learning rate. The clients objective then becomes
\begin{equation} \label{eq:fl-client-objective}
    F_k(w) =  \frac{1}{|D_i|} \sum_{b_{k,i} \in D_i} f_{b_k}(w)
\end{equation}
and so the shards objective will become
\begin{equation} \label{eq:fl-shard-objective}
    G_s(w) = \sum_k^K \frac{|D_k|}{D} F_k(w)
\end{equation}
This can be defined as weight updates by 
\begin{equation}
    w_s \leftarrow w_s + \sum_k^K \frac{|D_k| \Delta w_k}{D}
\end{equation}
This happens in parallel across all shards, where the global objective will become the sum of the results produced across all shards
\begin{equation} \label{eq:global-shard-objective}
    f(w) = \sum_s^S \frac{|D_s|}{D} G_s(w)
\end{equation}
We should note that this described approach follows the original FedAvg algorithm and can be modified in terms of fairness \cite{ref:q-fed-avg} among other averaging strategies.


\subsection{Shard Level Consensus} \label{sec:shard-consensus}
Each task deployed on the network can run an independent consensus mechanism. This allows for various task-related workloads to be accommodated by adapting the consensus algorithm. For example, when training large models where the performance of a single node may be limited, mechanisms such as PBFT \cite{ref:pbft} can be used. However, in shards with a fewer number of clients, we can make use of Raft consensus \cite{ref:raft-consensus}. This may benefit settings where model complexity is lower, and so a peer may be able to handle the throughput bottleneck of Raft consensus, determined by node performance.

Within the chosen consensus mechanism, we apply a pluggable policy to determine the acceptance of model writes. This policy can be substituted to handle alternative schemes (Section~\ref{sec:model-defence}). These schemes handle the identification of malicious clients by finding poisoned model updates or detecting Sybil attacks. It should be noted that while these plugins are applied at a task level to maintain the network's security, they can be upgraded with the smart contract that they are governing.

Another quality to consider is the detection of lazy clients. Detection of these clients may use a PN-sequence, which is applied to the model updates. This technique could then be applied within the shard level consensus to verify the clients' PN-sequence.

Because each shard contains a subset of the total participants for each task, the algorithmic complexity of consensus is significantly reduced to $\frac{\mathcal{C} \times \mathcal{P}_E}{\mathcal{S}^2}$. As a result, the network overhead needed to communicate model updates is also significantly reduced, which is the primary bottleneck in peer-to-peer federated learning. 

\subsection{Mainchain Consensus} \label{sec:mainchain-consensus}

\begin{figure}[!b]
	\centering
 	\includegraphics[width=\linewidth]{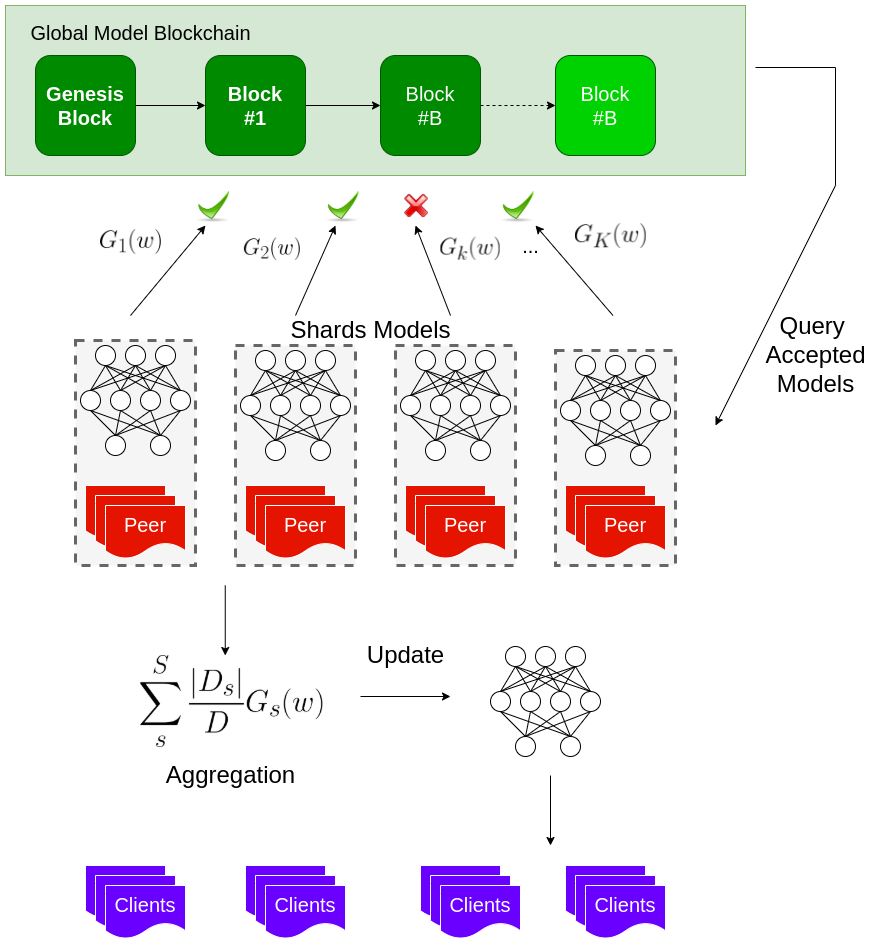}
	\caption{Aggregation of sharded models}
	\label{fig:global-aggregation}
\end{figure}

The mainchain is responsible for coordinating all verified aggregated model updates from each shard. This chain contains all participants across all shards; however, the activity on this chain is limited to shard level aggregation results and task proposals. The consensus for this follows the same pluggable consensus policy as the Shard level consensus (Section~\ref{sec:shard-consensus}); however, the submitting peers are limited only to the subset of endorsing peers on the shard chains. In this way, we limit the endorsing peers to those in possession of an aggregated model and are able to verify its authenticity. 

All accepted model updates to the mainchain will have been endorsed by each shard endorsing peers, which can safely be distributed and aggregated. If multiple models from a single shard have been accepted by disagreements between a shards endorsing peers, the model with more endorsements will win. So, if the endorsing peers submit the same model (model hashes are identical), the endorsing peers can safely evaluate the model update once. Figure~\ref{fig:global-aggregation} shows this process. The final step here is to post the final global model of the mainchain. This model represents the result after a single round of FL, and each of the shards will start the next round with these parameters. 

We note that a new round may begin earlier within a shard by assuming the model aggregated by a shard endorsing peers is valid before it is accepted to the mainchain. If this model is not accepted, then the round within that shard must be restarted. This should only happen if the majority of endorsing peers within a shard are compromised, and as such, the disruption will only affect the compromised shard.

\begin{figure*}[!t]
	\centering
 	\includegraphics[width=\linewidth]{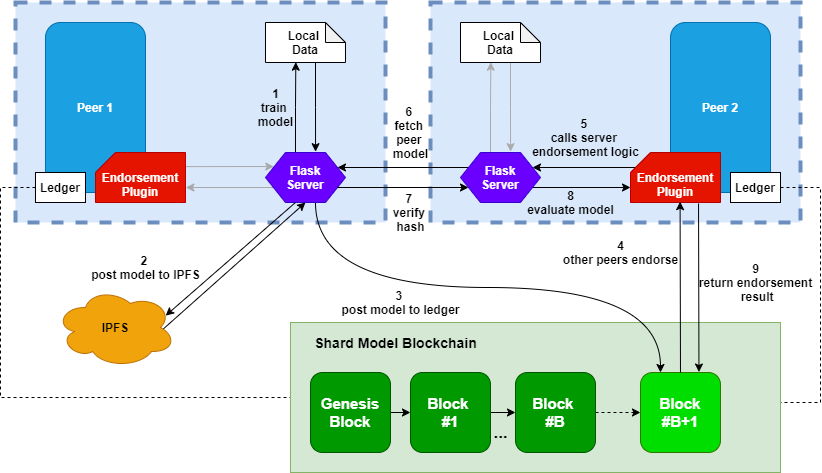}
	\caption{Transaction flow for shard level model submission}
	\label{fig:shard-level-model-submission}
\end{figure*}

\subsection{Workflow}
This section discusses the entire workflow of our proposed framework, discussing the steps required to produce a trained model.

Firstly, we define the participants' category in our system. We can break this into three categories: clients, peers, and endorsing peers, and described as follows.
\begin{enumerate}
    \item \textbf{Clients:} In a traditional FL setting, all participants are considered to be clients, where they are comprised of typical IoT devices such as smartphones, laptops, appliances, and vehicles, among others. Clients hold the private datasets that will be used for training models and are they are responsible for producing model updates. We consider the same assumption here, where clients do not validate models but rely on more powerful devices to ensure the system's security. This is the lightest class of participants in terms of resource consumption.
    \item \textbf{Peers:} Peers are responsible for holding copies of the ledger. From this pool of peers, a committee can be elected to reduce the evaluation overhead. This helps take the load off of endorsing peers and allows peers to offer API services for clients.
    \item \textbf{Endorsing peers:} Members of the committee are responsible for evaluating model updates during consensus. They hold their own private dataset to endorse other model updates and can submit their own model updates, although they are not required to do so. Furthermore, the policy for evaluation of model weights can be modified, and these peers additionally must check for valid authentication of the write-set. Finally, committee members must participate in the mainchain consensus to determine which shard level updates will be accepted.
\end{enumerate}
These participants will then participate in one or more tasks, as managed by the mainchain smart contract. Starting from a fresh network, we will describe the following process.

\subsubsection{Task Proposal} The procedure begins with a task proposal on the application layer. It will be proposed on the mainchain in the form of a smart contract, and it outlines the requirements and task description. We leave the details of this specification open, as we may not know the complete set of features or optimization level objectives as such in a vertical FL (VFL) scenario \cite{ref:federated-learning-concepts}. Once this proposal has garnered enough interest in client registration, a new shard will be provisioned, and shard-related smart contracts will be deployed to the newly created shard(s). These processes can be thought of as an open request for participants to collaborate on a specific task.

\subsubsection{Client Training}

Within a specific shard, an off-chain coordination network will be formed as defined by the proposal. This could utilize existing FL frameworks and best practices, and new paradigms. As such, the fraction of clients fit within a specific round will be determined off-chain by this process and exclude, for example, currently unavailable clients due to battery or network coverage issues. These selected clients are then responsible for producing a model update to the global model weights $w$. Each required client will then compute their local updates as seen in Eq.~\eqref{eq:fl-client-objective}. We can see this as Step~1 in Figure~\ref{fig:shard-level-model-submission}.

\subsubsection{Off-Chain Model Storage}
A client will then optionally upload a model to an off-chain cache such as InterPlanetary File System (IPFS) \cite{ref:ipfs}. At this stage, the model can be downloaded publicly; however, it has not yet been verified nor recorded on the ledger yet. This step is used to record historical models or take the network load off of serving local model updates to peers. We can see this as Step~2 in Figure~\ref{fig:shard-level-model-submission}.

\subsubsection{Model Submission}
Once the model has been uploaded and is available for download, the client will submit the update metadata by invocation of the shard level smart contract. This update should include information such as the model's hash value and a link from which the model can be downloaded. This can be seen in Step~3 in Figure~\ref{fig:shard-level-model-submission}.

\subsubsection{Peer Endorsement}
The peers on the shard must endorse this model update according to the shard level consensus mechanism (Section~\ref{sec:shard-consensus}). This work could be offloaded to the peers' local workers, where additional computing resources may be present. This can be seen in Steps~4 and 5 in Figure~\ref{fig:shard-level-model-submission}.

\subsubsection{Model Evaluation}
The peers' local workers request the model weight update from the link specified in the model update. The downloaded model will then be verified against the submitted hash to ensure integrity. Once the model is verified, it can be evaluated by various techniques; for example, the peer may test the model against its local dataset and reject the update on sufficient degradation in performance, such as in the RONI defence. This would be ideal in cases where the data distribution is known to be IID. Alternative solutions such as the FoolsGold scheme may be used for non-IID data, which could be used to identify poisoning attacks from Sybils. The results of this will be returned from the worker back to the peer. Here, we have summarized Steps~6-8 in Figure~\ref{fig:shard-level-model-submission}.

\subsubsection{Shard Aggregation}
Endorsed model updates are written to the ledger according to the consensus quorum. The off-chain FL process can continue by aggregating the models from the current round if they have been accepted into the ledger. These models will then be downloaded and aggregated by the coordinating servers according to Eq.~\eqref{eq:global-shard-objective}. These aggregated models will then be published to the mainchain. Similarly, these aggregated shard updates are only be accepted if the committee of endorsing peers reaches a consensus on the update governed by the mainchains consensus mechanism (Section~\ref{sec:mainchain-consensus}).

\subsubsection{Global Aggregation}
Finally, all accepted shard level model updates accepted to the mainchain will be downloaded and aggregated by the shard level servers. The peers can easily verify these models by checking the hash of the new distributed global model against the posted hash from the mainchain for the finalized FL round.

\section{Experiments}\label{sec4}
To demonstrate the effectiveness of our proposed sharding solution, we evaluate the performance in a Proof of Concept (PoC) implementation\footnote{https://github.com/blockchain-systems/ScaleSFL}. The purpose of experimentation here is first to measure the throughput of a FL workload scale with respect to the number of shards. Secondly, we measure the sharding impact on model performance in both IID and non-IID scenarios. It is important to ensure the quality of models and keep the rate of convergence low when a potential subset of peers are malicious.

To run these experiments, we use Hyperledger Fabric \cite{ref:fabric}, a permissioned distributed ledger platform. This platform offers a modular architecture allowing for plug-n-play consensus and membership services. Hyperledger Fabric offers an execute-order-validate architecture for transactions, allowing each peer to evaluate models in parallel within each shard. This is opposed to most public blockchain platforms that first order transactions and execute them sequentially. Fabric additionally offers modular endorsement and validation policies which we will use to implement custom consensus logic within an endorsement plugin. We will read in a proposal response payload, sending the given transaction read-write set to the peers' worker for validation. This read-write set contains the necessary information to endorse or reject the proposed model update. This will be built directly into the Fabric peers. 

Our solution workflow may be adapted to alternative permissioned platforms such as Quorum \cite{ref:quorum}, or public platforms such as Ethereum \cite{ref:ethereum}, which may allow for additional reward allocation schemes. Implementations on these platforms may consider the use of child chains utilizing existing layer-2 scaling solutions. We choose Hyperledger Fabric for our PoC primarily for the flexibility of consensus and membership services during development.

\begin{table*}
	\caption{Experimental Configuration}\label{tab:experimental-configuration}
	\begin{tabular}{cccccc}
		\toprule
		Component&Type&CPU&GPU&RAM&Disk (SSD)\\
		\midrule
		Caliper Benchmark&Caliper 0.4.2&Ryzen 7 3800X@3.9 GHz&NVIDIA RTX 3080&64GB&1TB\\
		Fabric Peer (Shard)&Fabric 2.3.3&Ryzen 7 3800X@3.9 GHz&NVIDIA RTX 3080&64GB&1TB\\
		Fabric Peer Worker&Python 3.9, PyTorch 1.10.1&Ryzen 7 3800X@3.9 GHz&NVIDIA RTX 3080&64GB&1TB\\
		\bottomrule
	\end{tabular}
\end{table*}

Tests here are run locally on a single machine simulating a Fabric test-network with a single orderer running Raft \cite{ref:raft-consensus}, and eight peers, each with their own organization, along with a certificate authority. Each peer worker runs on a single thread, allowing each peer to act independently. The experimental configuration can be seen in Table~\ref{tab:experimental-configuration}.

We implement two smart contracts to implement both the mainchain consensus and shard level consensus mechanisms. Smart contracts, otherwise known as chaincode in Fabric, will be deployed to a specific channel within Fabric. We use channels to simulate shards, where each channel operates independently with the ability for different membership and endorsement policies per channel. Additionally, we relax the previous definitions of endorsing peers, such that $P\colon \{Peers\} = P_E\colon \{Endorsing~peers\}$. We deploy a shard level smart contract we term the ``models" chaincode to each channel. Each participant operating within a shard will be assigned to a shard by our managing contract. We deploy the mainchain consensus on its own channel, in which every peer from every shard will join. We term this the catalyst contract, where it is responsible for aggregating a global model from each of the shards. As mentioned previously, model updates are stored off-chain. In this implementation, each peers worker hosts the models locally with a gRPC server.

We coordinate the FL processes off-chain while pinning updates back to each shard's ledger. This coordination is abstracted from the approach discussed previously; in this case, we use the Flower \cite{ref:flower} an open-source extensible FL framework. In this implementation, during aggregation, submitted models are verified to have been endorsed, which can be checked by querying the peers' local ledger to reduce a round of communication. This is conducted by creating a custom strategy within the Flower server and modifying the aggregated fit to filter out any updates which are not present on-chain, by querying the models' smart contract. We additionally implement differential privacy using Opacus \cite{ref:opacus}, applied during client training. We use an $(\epsilon, \delta)$ target of (5, $1\mathrm{e}{-5}$), a noise multiplier of 0.4, and a max gradient norm of 1.2.

\subsection{Evaluation}
For evaluation, we employ Hyperledger Caliper \cite{ref:hyperledger-caliper}, an open source benchmarking tool allowing for various workloads to be conducted for our system under test (SUT). These tests are performed by an independent process that generates transactions to be sent to the SUT by a specified configuration, monitoring for responses to determine metrics such as latency, throughput and success rate. We conduct tests with varying numbers of workers, transactions counts, and transactions per second (TPS) to evaluate the limits of the system.

We run several workloads to test the effects of sharding on transaction throughput and model performance. The benchmark tasks include:
\begin{itemize} 
    \item \textbf{Update creation throughput:} We first run an experiment to test the throughput of model creation. To do this, we generate a number of models updates, make the parameters available locally, and have the endorsing peers evaluate them during consensus. In this way, we can measure how sharding affects the performance of the consensus mechanism by parallelizing the workload among shards. We aim to verify our initial claim that the global computations required with become $\frac{\mathcal{P} \times \mathcal{C}}{\mathcal{S}}$.
    \item \textbf{Model performance:} We evaluate the performance of the models across two scales: (1) the global number of epochs computed and (2) the number of gradients, which provide more accurate measures of performance for decentralized training \cite{ref:bfl-shard}. While we are primarily concerned with the performance of integrating existing FL solutions into the blockchain, sharding should allow a larger fraction of clients to be fit, allowing for model performance to be increased.
\end{itemize}

\subsection{Datasets}
For experimentation, we use three datasets, namely the MNIST dataset \cite{ref:mnist} consisting of 60,000 28x28 images of handwritten digits, and the CIFAR-10 dataset \cite{ref:cifar10} which consists of a dataset of 60,000 32$\times$32 images separated into 10 classes with 6,000 images per class. We use these to represent IID data, where these classes of data are split evenly between clients. 

For non-IID scenarios, we use the LEAF dataset \cite{ref:leaf-dataset} which can be used to test model performance by partitioning the data between peers in a reproducible way. In this setting, the classes of data may not be evenly split. For example, in the case of the Federated Extended MNIST (FEMNIST) dataset, characters were split by the writer of the digit or character. This can be a strong benchmark to model real-world distributions of data.

These datasets are used as a common baseline for comparison and evaluation against previous solutions \cite{ref:federated-learning, ref:bfl-async, ref:bfl-shard}, showing the addition of a sharding solution maintains competitive results on model performance. While a non-IID assignment of data across shards may influence the generated global model during the global aggregation stage, the selection of mainchain defence mechanisms plays the largest role in model convergence. Thus, the results generated by the chosen datasets demonstrate the applicability of sharding solutions to existing Blockchain-based methods.

\subsection{Results}
We created multiple workloads for the \textit{Update Creation Throughput} benchmark to explore the effectiveness of our sharding solution. These workloads are conducted without malicious updates using the MNIST \cite{ref:mnist} dataset to evaluate the performance of model creation transactions. We set the timeout period for each transaction to 30 seconds, allowing us to consider the number of failed transactions as stale (i.e., not malicious). In this experiment, each of the clients evaluated the update against their entire local dataset, in which we have used the entire test split of the MNIST dataset for each client. 

\begin{figure}[!t]
	\centering
	\includegraphics[width=\linewidth]{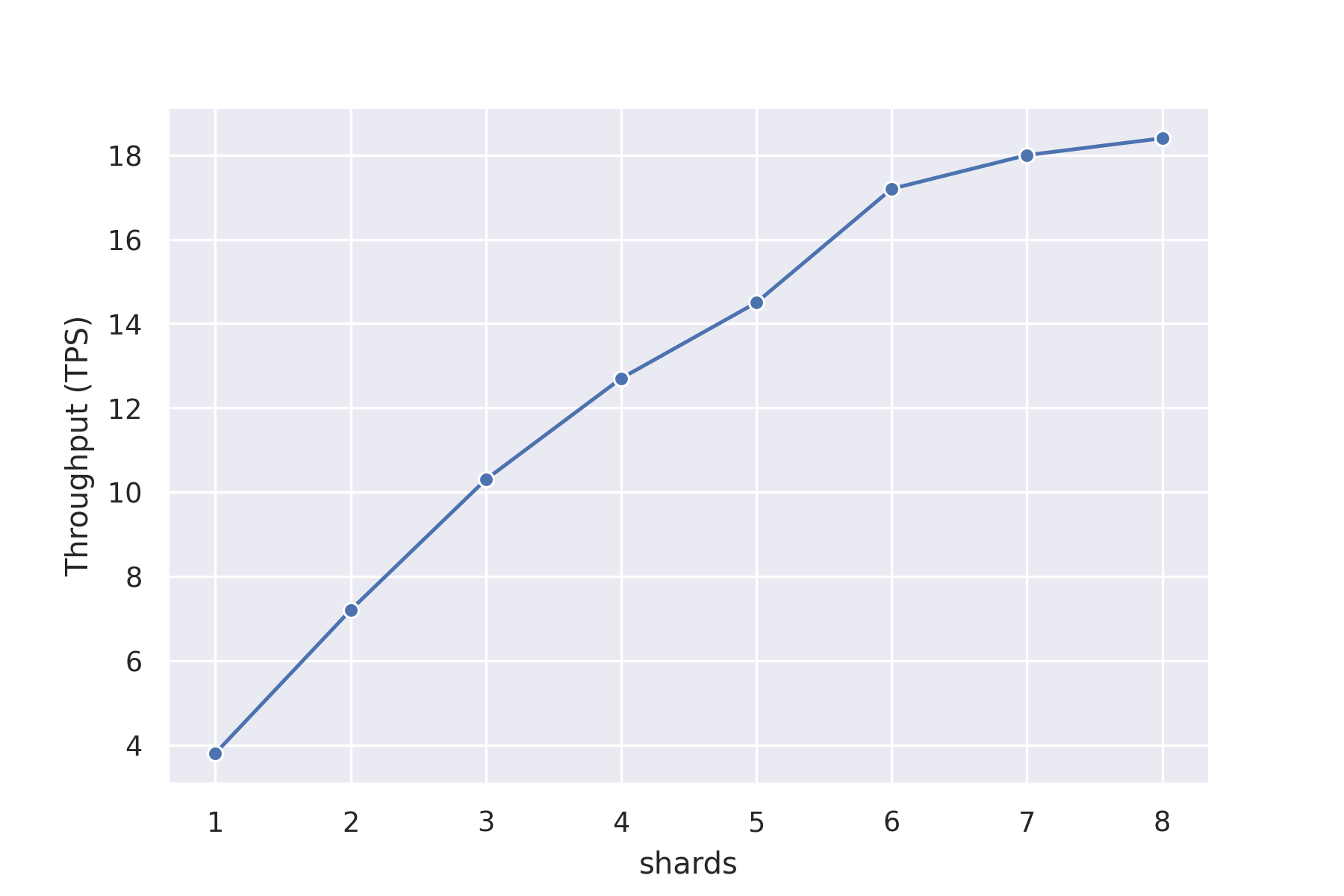}
	\caption{\#shards vs. system throughput (TPS)}
	\label{fig:benchmark-shards-by-throughput}
\end{figure}

We first measure the throughput with respect to the number of shards, as seen in Figure~\ref{fig:benchmark-shards-by-throughput}. This workload uses two Caliper workers, evaluated over 200 transactions, where we can see the throughput for each independently operating shard which allows the global throughput to scale linearly. This directly reflects the time for the evaluation of a model that is the primary bottleneck. We note here that the sent TPS for each number of shards is set just above its throughput in order to saturate the system for the evaluation. In this workload, we used the same number of sent transactions for consistency; however, higher numbers of shards have shown improvements in throughput for increased transactions sent, while a lower number of shards may begin to fail earlier.

\begin{figure}[!t]
	\centering
	\includegraphics[width=\linewidth]{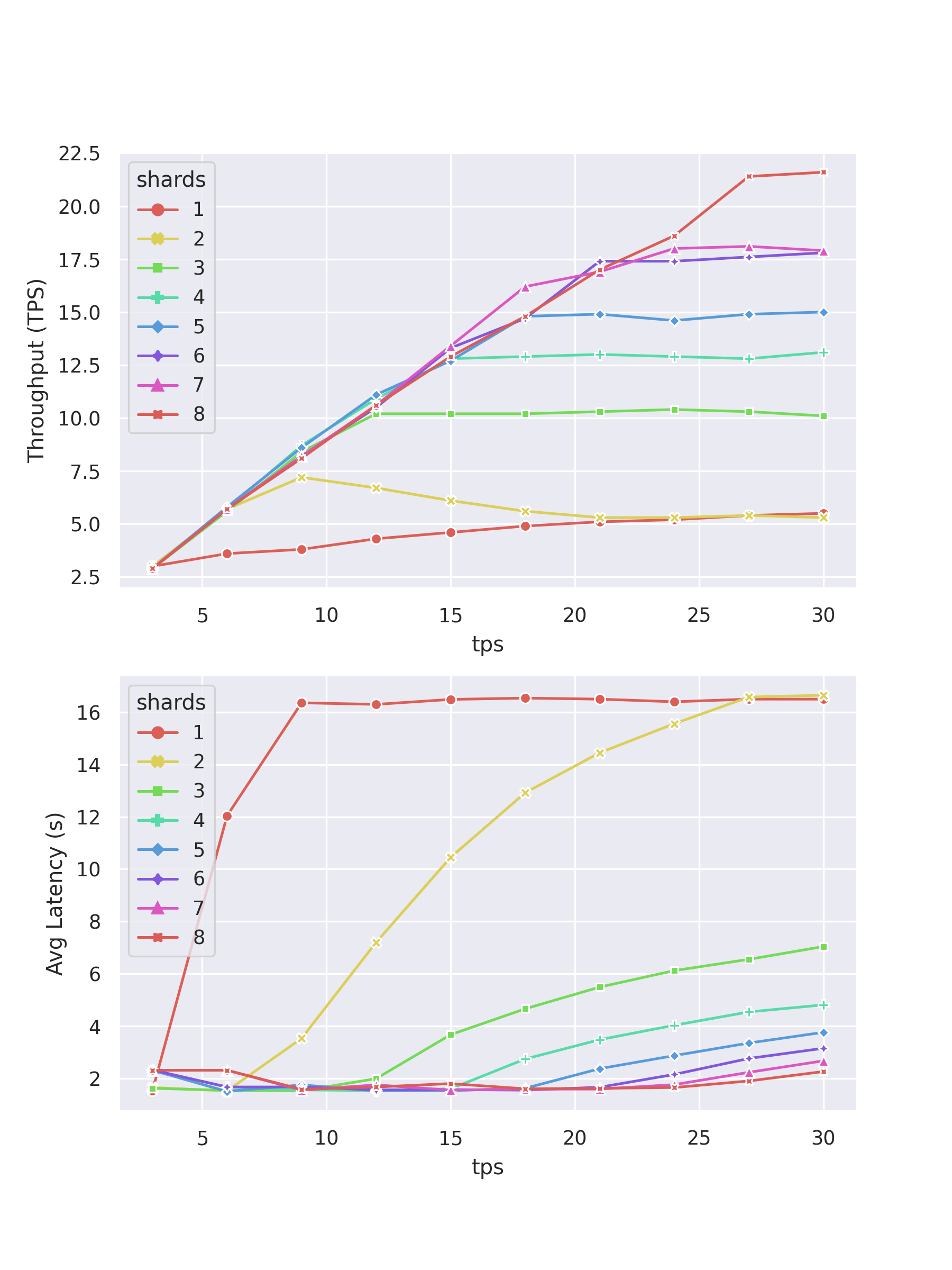}
	\caption{Sent TPS vs. system throughput (TPS) \& average response latency}
	\label{fig:benchmark-tps-by-throughput-avglatency}
\end{figure}

To test the maximum throughput achieved by our system, we measure the sent TPS against the system throughput and average latency. See Figure~\ref{fig:benchmark-tps-by-throughput-avglatency}. This workload is run with 2 caliper workers over 200 transactions. As we increase the sent TPS, the system will reach a point where it becomes saturated and is unable to handle a higher TPS. We can see this by looking at the point where the average latency begins to increase in the figure, where the throughput is maximized. We conduct this workload by measuring sent TPS in increments of 3, starting from 3 TPS. We can see each additional shard increases the overall system throughput similarly in Figure~\ref{fig:benchmark-shards-by-throughput} while demonstrating the maximum throughput limit.

\begin{figure}[!t]
	\centering
	\includegraphics[width=\linewidth]{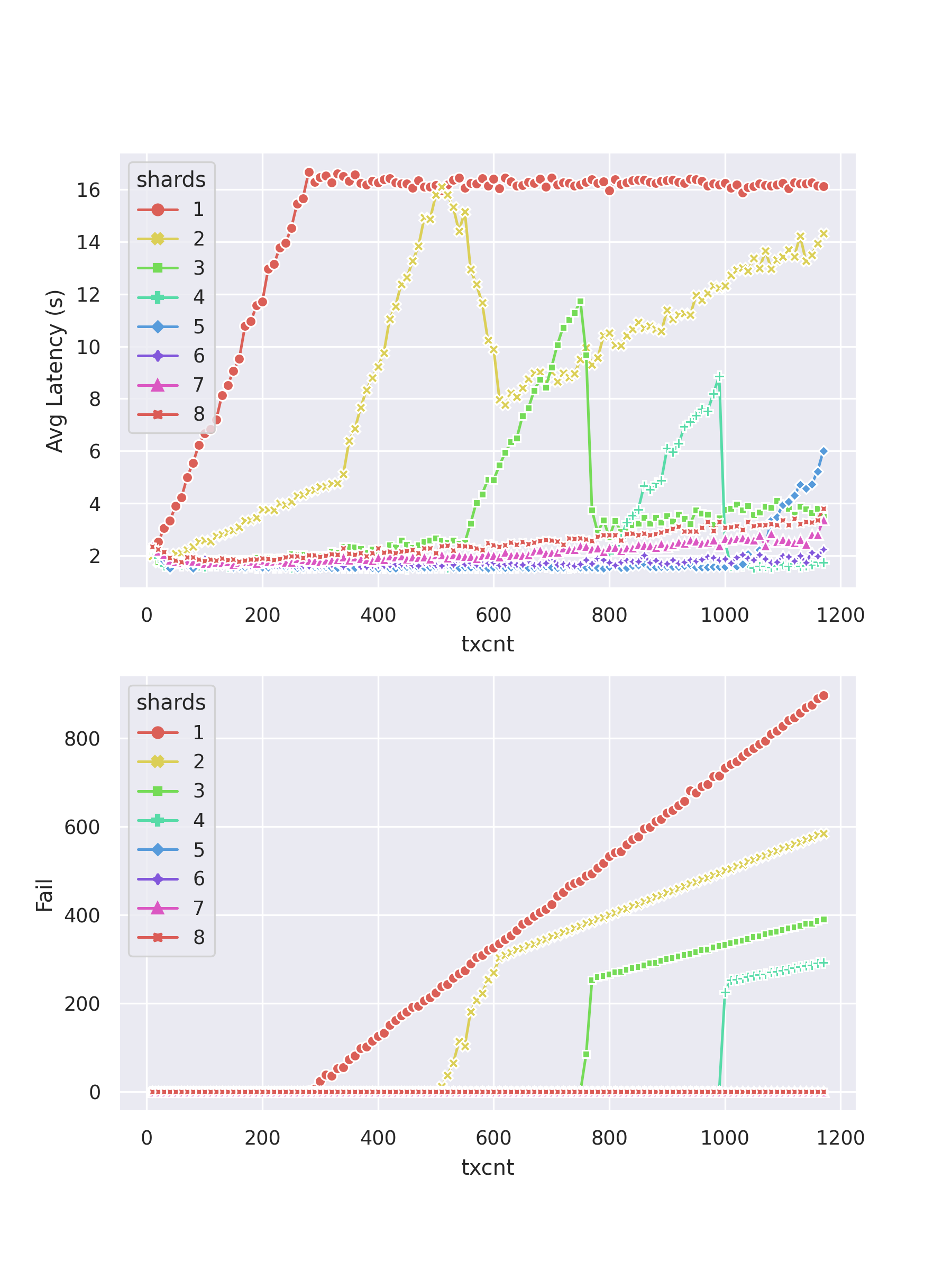}
	\caption{Transaction count sent vs. average response latency \& failure count}
	\label{fig:benchmark-txcnt-by-fail-avglatency}
\end{figure}

To explore the limits of a usage surge, we measure the number of transactions sent by system throughput, average latency, and failure rate. In this benchmark, we are interested in the system's behaviour under a workload with a higher sent TPS than throughput. This workload is run with 2 caliper workers and a sent TPS just above the maximum throughput as seen in the previous workload. See Figure~\ref{fig:benchmark-txcnt-by-fail-avglatency}. When the system cannot handle the number of transactions currently queued, the average latency will spike, and the number of failed requests will increase because the system begins timing out requests. The system will then go through a ``flush" period where many long-running requests begin to fail, so the average latency of successful requests will drop. We note here that the average latency peaks at around 16 seconds due to the average between the maximum latency (the timeout limit) and the minimum latency, which corresponds to the time it takes to evaluate a model fully.

\begin{figure}[!t]
	\centering
	\includegraphics[width=\linewidth]{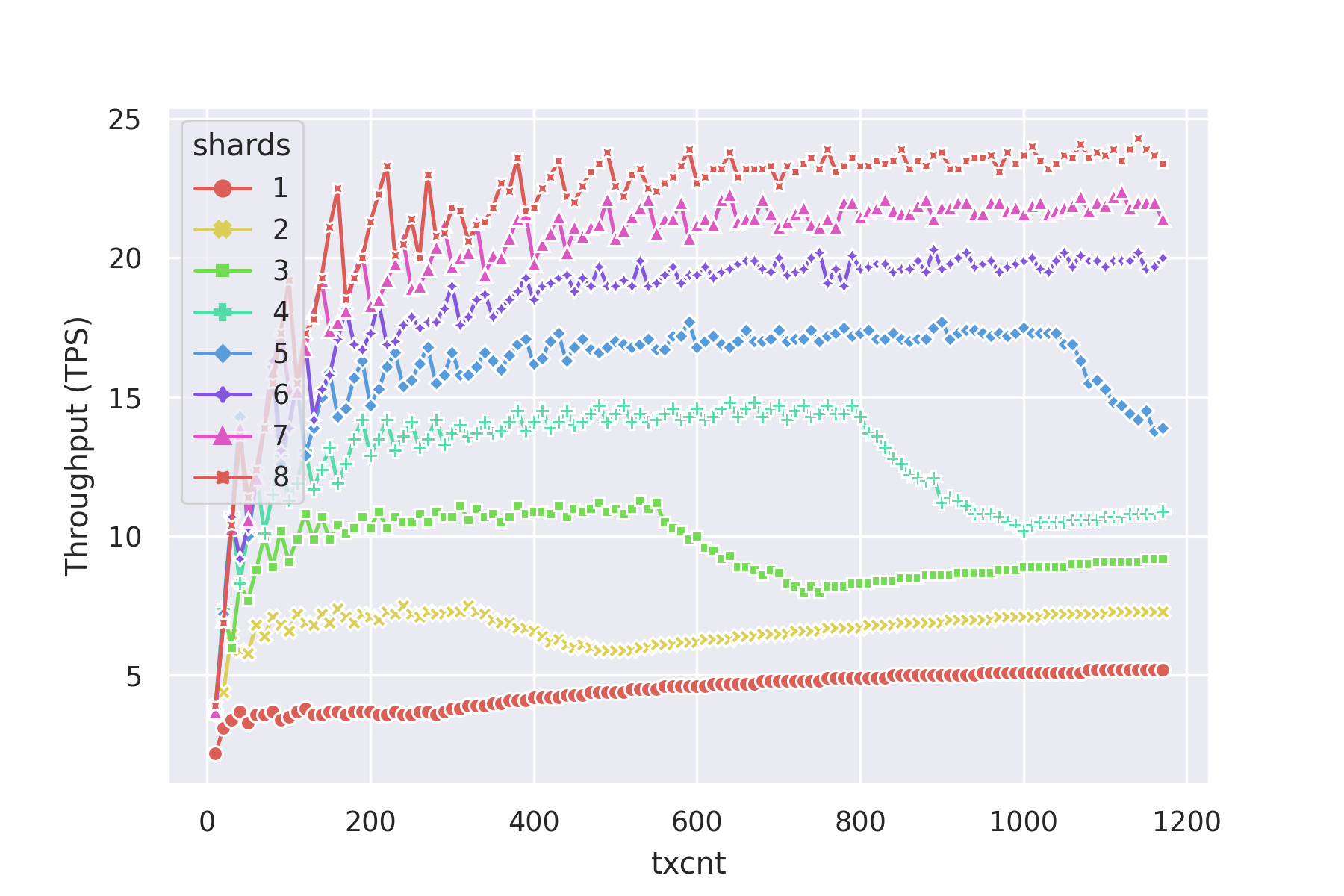}
	\caption{Transaction count sent vs. System throughput (TPS)}
	\label{fig:benchmark-txcnt-by-throughput}
\end{figure}

Figure~\ref{fig:benchmark-txcnt-by-throughput} shows the throughput over this workload, where the throughput of the system decreases when it is no longer able to handle the number of incoming requests, as the overhead to process these additional requests limits the number of successful requests processed. The decrease in throughput corresponds to the increase in average latency previously observed.

\begin{figure}[!t]
	\centering
	\includegraphics[width=\linewidth]{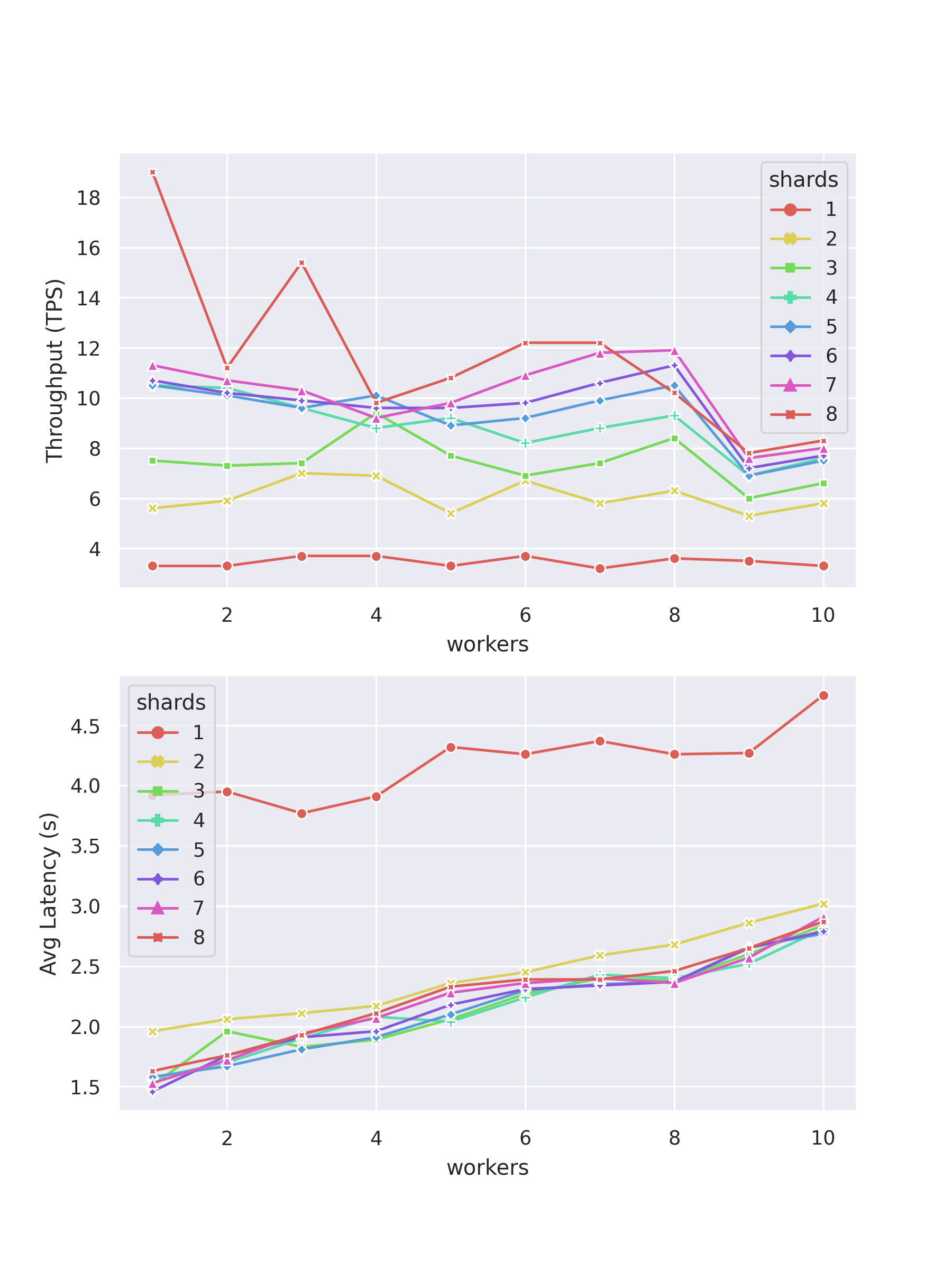}
	\caption{Caliper workers vs. system throughput (TPS) \& average response latency}
	\label{fig:benchmark-workers-by-throughput-avglatency}
\end{figure}

We finally test how the system handles concurrent requests. To do so, we run multiple workloads, varying the number of caliper workers, and measuring the system throughput and average latency. This workload configuration sends 200 transactions, with the sent TPS equal to the previously-mentioned maximum throughput. This can be in Figure~\ref{fig:benchmark-workers-by-throughput-avglatency}. We observe that the throughput of the system with respect to the number of workers is quite noisy, however, there is a general downward trend in the system throughput. Increasing the number of caliper workers here allows us to scale the workload generation, however since the endorsement workers evaluating the model operate sequentially in these experiments (limited to a single thread), The throughput is limited. Similarly, we can see an upward trend in average latency caused by each transaction's time waiting for execution in the queue. We can see that the number of shards plays the largest factor here as workloads with more than 2 shards are tightly grouped with respect to average latency due to these workloads being able to operate in parallel across shards. 

\begin{figure}[!t]
	\centering
	\includegraphics[width=\linewidth]{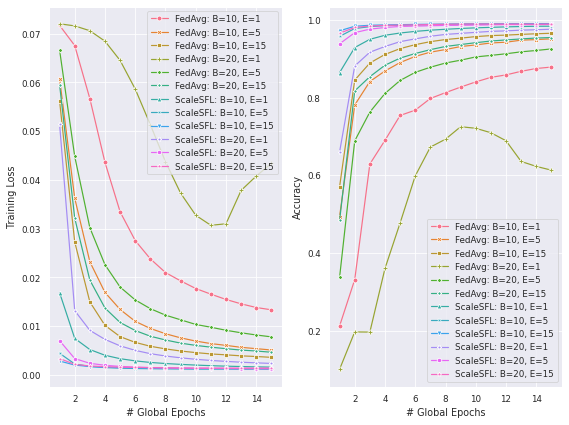}
	\caption{Training loss and testing accuracy of CNN model on the MNIST dataset}
	\label{fig:benchmark-model-performance}
\end{figure}

\begin{table}
	\caption{Best accuracy with varying minibatch size (B) and number of local epochs (E)}\label{tab:benchmark-model-performance}
	\begin{tabular}{cccccc}
		\toprule
		$B$&$E$&FedAvg (Accuracy)&ScaleSFL (Accuracy)\\
		\midrule
		10&1&00.8784&0.9835 \\
		10&5&0.9511&0.9897 \\
		10&15&0.9654&0.9896 \\ \hline
		20&1&00.7249&0.9758 \\
		20&5&0.9251&0.9881 \\
		20&15&0.9544&0.9889 \\
		\bottomrule
	\end{tabular}
\end{table}

To measure \textit{model performance}, we measure the effect of various training parameters. We use minibatch sizes of $B \in \{10, 20\}$, and local epoch sizes of $E \in \{1, 5, 15\}$. We present results here from the non-IID MNIST dataset, however similar results hold for both CIFAR-10, and LEAF benchmarks. The effects of both minibatch size and local epoch sizes can be seen in Figure~\ref{fig:benchmark-model-performance}, and Table~\ref{tab:benchmark-model-performance}, where we compare our ScaleSFL solution and the traditional FedAvg algorithm. This comparison is performed under the assumption all clients are honest, and thus we are simply comparing the effect of sharding. Both training loss and accuracy are considered, and workflows are run for 15 global epochs, where the number of ScaleSFL shards is set to 8, with each shard sampling eight clients each round, making for a total of 64 clients. The FedAvg algorithm is comparably run with 64 clients, with both methods using a client learning rate of $\eta_k = 1e-2$.

In Figure~\ref{fig:benchmark-model-performance}, we observe the convergence rate of ScaleSFL is faster than the FedAvg algorithm, generally reaching an accuracy of $0.98$ within the 15 global epochs considered. This is due to the parallelized nature of ScaleSFL, where each shard can simultaneously filter and produce viable updates with reasonable client sampling rates, combining them to produce a stronger global model at each step.

\section{Discussion}\label{sec5}
\paragraph{\textbf{Hierarchical Sharding}}
Since each shard operates independently, several additional improvements over a traditional FL workflow can be achieved. For example, a typical device selection step is usually performed, sampling a subset of the available devices for each round. However, since the population of each shard is much smaller than the global population, each shard's ability to update and maintain active devices becomes more efficient, and during aggregation, a much larger sampling of devices can be used from the global population. 

Additional optimizations could be achieved by placement of clients based on region \cite{ref:hierfavg}. Since each client may submit their model to an off-chain cache such as IPFS, shards with a population based within the same region could reduce overhead latency. In this way, the overhead only occurs during the global aggregation phase. The possibility of single-shard takeover attacks may be introduced by not using a random sampling scheme. However, as noted before, this disruption will only affect the compromised shard since all endorsing peers from all shards must agree on which shard level updates are aggregated. This may allow for region based sampling algorithms to be used for the assignment of participants. These strategies should be chosen depending on the underlying task, potentially chosen by the task submitter, or consensus among task participants prior to shard creation. 

Alternative participant sampling can be considered in the case of cross-silo or consortium based settings such as medical or financial organizations. For example, it may be beneficial to allow clients of a particular organization to be grouped under a single shard \cite{ref:bfl-shard}. Fine-grained control oversampling and permissions are delegated independently within each shard.

\paragraph{\textbf{Model Provenance}}
By pinning model information back to the mainchain, we provide an open model hub for sharing and distributing trained models in a decentralized fashion. Model submission to IPFS additionally allows for transparent model check-pointing and disaster recovery. In the case that a bug has been introduced at some point in the FL process, either intentionally through malicious attack, or by data bias or incompatibility, previous model checkpoints may be restored, and a new task may be initiated using this saved model checkpoint.

\paragraph{\textbf{Rewards Allocation}}
While our implementation of ScaleSFL uses Hyperledger Fabric, alternative implementations may consider platforms such as Ethereum \cite{ref:ethereum} for implementation of reward mechanisms. Since training models in a permissionless fashion may result in decreased contributions, due to the availability of models through a model hub, and without expending computing resources, rewards become an integral part of the workflow. While access to the models should remain free, contributing computing resources should be rewarded by distributing rewards to the contributing clients. Additional incentives could be provided by task contributors or interested clients, to ``sweeten the pot", encouraging community contributions. Additionally, with the incorporation of a native cryptocurrency, submitting models transactions could incur a small gas fee. This would prevent attempted Denial of Service (DOS) attacks against the system since the rewards for model contributions are only realized for non-malicious updates. However, common DOS attacks may generate random model updates or send updates with dead cache links.

\paragraph{\textbf{Alternative Attacks}}
Up to this point, we have discussed malicious clients as clients which send poisoned model updates. However, malicious clients may attempt various alternate types of attacks, such as uploading a different model than specified in the task request or making the model update very large, attempting to perform a DOS attack on a shard. These attacks may be prevented by checking the size of the model uploaded to the cache prior to download. However, a more clever client may attempt to submit a legitimate model many times, attempting to either reap the rewards for training, or to perform a DOS attack against the system. These clients are known as lazy nodes \cite{ref:fl-meets-bc}. They additionally attempt to copy another client's model update during a gossip phase. This problem has severe negative effects on global model performance, so updates from these lazy nodes should be discarded. One way to recognize these lazy nodes is to publish model updates with a pseudo-noise (PN) sequence, and then subsequently publish the PN sequence after each client has published their model update. Lazy nodes will be spotted by checking the correlation between the model updates and the published PN sequence \cite{ref:fl-meets-bc}. Gas fees should deter spotted clients, and Sybils from attempting to submit further model updates. Incorporating the defence into our workflow would modify the off-chain model storage step, ensuring all models submitted during model submission have a valid PN signature.

\section{Conclusions}\label{sec6}
In this paper, we proposed a sharding solution for blockchain based federated learning. We described a modular workflow, allowing for pluggable defence mechanisms and poisoning mitigations strategies. This workflow supports off-chain FL frameworks to be used, while pinning verified model updates to the blockchain. We additionally implement a proof-of-concept of ScaleSFL using Hyperledger fabric. Our experimental results demonstrated that ScaleSFL improves validation performance linearly with the addition of shards. This helps address the prevalent scalability issue related to blockchain consensus by demonstrating the impact of sharding.

As {\em ongoing and future work}, we further explore the details of a sharded solution applied to the blockchain.
Specifically, we aim to simulate malicious attacks on the system via model poisoning updates. 
This would show the effectiveness of a blockchain solution to provide security against bad actors. 
Additionally, we will explore the application of a reward mechanism for participating in the model training process for our current system. This could be explored on other chains such as Ethereum \cite{ref:ethereum}, where the use of native tokens is well-supported. Finally, we would like to implement a more full-featured implementation of this PoC, including dynamic shard creation and allowing model proposition through our catalyst contract.

\begin{acks}
This work is partially supported by NSERC (Canada) and University of Manitoba.
\end{acks}

\bibliographystyle{unsrt}
\bibliography{report}



\end{document}